# Enhancing condensation on soft substrates through bulk lubricant infusion

Chander Shekhar Sharma[2], Athanasios Milionis[1], Abhinav Naga[4], Cheuk Wing Edmond Lam[1], Gabriel Rodriguez[1], Marco Francesco Del Ponte[1], Valentina Negri[1], Hopf Raoul[3], Maria D'Acunzi[4], Hans-Jürgen Butt[4], Doris Volmer[4], Dimos Poulikakos[1†]

[1]Laboratory of Thermodynamics in Emerging Technologies, Department of Mechanical and Process Engineering, ETH Zurich, 8092 Zurich, Switzerland

[2]Thermofluidics Research Lab, Department of Mechanical Engineering, Indian Institute of Technology Ropar, Rupnagar, 140 001 Punjab, India

[3] Institute of Mechanical Systems, Department of Mechanical and Process Engineering, ETH Zurich, 8092 Zurich, Switzerland

[4]Max Planck Institute for Polymer Research, Ackermannweg 10, D-55128, Mainz, Germany

[†]E-mail: dpoulikakos@ethz.ch. Phone: +41 44 632 27 38. Fax: +41 44 632 11 76

**Abstract**

Soft substrates such as polydimethylsiloxane (PDMS) enhance droplet nucleation during the condensation of water vapour, because their deformability inherently reduces the energetic threshold for heterogeneous nucleation relative to rigid substrates. However, this enhanced droplet nucleation is counteracted later in the condensation cycle, when the viscoelastic dissipation inhibits condensate droplet shedding from the substrate. Here, we show that bulk lubricant infusion in the soft substrate is a potential pathway for overcoming this limitation. We demonstrate that even 5% bulk lubricant infusion in PDMS reduces viscoelastic dissipation in the substrate by more than 30 times and more than doubles the droplet nucleation density. We correlate the droplet nucleation and growth rate with the material properties controlled by design, i.e. the fraction and composition of uncrosslinked chains, shear modulus, and viscoelastic dissipation. Through *in-situ*, microscale condensation on the substrates, we show that the increase in nucleation density and reduction in pre-coalescence droplet growth rate is insensitive to the percentage of lubricant in PDMS. Our results indicate the presence of a lubricant layer on the substrate surface that cloaks the growing condensate droplets. We visualize the cloaking effect and show that lubricant infusion in PDMS significantly increases the rate of cloaking compared to PDMS without any lubricant infusion. Finally, we show that the overall enhanced condensation due to bulk lubricant infusion in PDMS leads to more than 40% increase in dewing on the substrate.

# 1. Introduction

Condensation of water vapor on a colder surface is a widely observed phenomenon and manifests itself visibly to the human eye, for example, as fogged windscreens or glasses on a humid day. In industry, condensation is critical in multiple applications such as power generation (1), water desalination (2), and dew water harvesting (3, 4) and thermal management (5). Tuning and predicting the outcome of condensation is challenging because of the interplay of several processes. It is well established that condensation of a fluid on rigid substrates proceeds in four distinct steps – formation of initial condensate nuclei, growth of individual droplets through direct condensation, droplet coalescence, and eventual removal of the condensate from the surface (6). On hydrophilic substrates, coalescence of the individual droplets usually results in a continuous condensate film. However, efficient condensation requires that the fluid condenses as distinct droplets that rapidly shed from the surface (7). This condensation mode can be realized by hydrophobizing the substrate through the modification of the surface roughness and chemistry. Depending on the degree of hydrophobicity and morphology of the surface texture, condensate droplets can be shed by multiple mechanisms such as gravity-induced depinning (7), coalescence induced droplet jumping (8–10), cascading coalescence (4), and Laplace pressure gradients (11). However, this enhanced condensate droplet shedding on rigid hydrophobic substrates is achieved at the cost of reduced nucleation density compared to hydrophilic substrates, which is a critical limitation for water harvesting applications (12, 13).

Soft solid substrates such as Polydimethylsiloxane (PDMS) overcome the limitation concerning nucleation density by lowering the energetic threshold for heterogeneous nucleation. This is because the condensate droplets can trigger substrate deformation through elasto-capillary effects and reduce the overall energy of the system consisting of substrate-water and water-water vapor interfaces (14, 15). Unfortunately, this increase in nucleation density is achieved at the cost of reduced droplet mobility due to the formation of a distinct wetting ridge around the droplet (16). As the droplet slides, this wetting ridge causes viscoelastic dissipation within the substrate, otherwise known as viscoelastic braking (17). This effect can be far more significant than the viscous dissipation within the liquid condensate droplets, thereby making the material properties of the soft substrate the governing factor behind the inhibited motion of the droplets (16, 18).

The above-described characteristics of high condensate nucleation density but inhibited droplet movement on soft substrates highlight a quandary for practical applications of soft substrates in condensation applications, specifically water harvesting from humid air. In recent years, slippery, lubricant-infused rigid textures have been reported, potentially addressing the dilemma mentioned above (19, 20). Such a hybrid texture can present enhanced nucleation density and rapid shedding of condensate droplets due to low contact angle hysteresis (21–23). However, lubricant-infused textures may be limited by the gradual draining of the thin lubricant layer from the micro and nanometric surface features (22, 24).

Here, we design a PDMS-based organogel to achieve high nucleation density while maintaining a high rate of droplet shedding. Our approach involves increasing the fraction of uncrosslinked chains within the substrate through the bulk infusion of a lubricant. The resulting organogel (25) allows controlled reduction in viscoelastic dissipation in PDMS samples while retaining the inherently high droplet nucleation density. We perform careful material characterization of the substrates and *in-situ* micro- and macroscale investigations of condensation. We find that this adjustment of the proportion of uncrosslinked chains directly influences all the three key processes in the heterogeneous dropwise condensation cycle - namely droplet nucleation, droplet growth, and droplet shedding from the surface. The uncrosslinked chains slowly diffuse outwards, forming a layer of lubrication. This lubricant layer enhances droplet movement on the surface (26–29) and increases droplet nucleation during condensation. However, it also results in the cloaking of condensed drops, thereby lowering the individual microdroplet growth rate. Overall, however, the enhancement in droplet nucleation and sliding more than compensate for any reduction in droplet growth rate, resulting in a significant enhancement in water condensation rate. We provide experimental evidence for this enhancement in terms of condensate water collection. PDMS substrates with bulk lubricant infusion condense more than 40% more water than conventional PDMS substrate under similar dewing conditions. We also systematically characterize the various material properties that affect the overall condensation process — material stiffness characterization through bulk stiffness and wetting ridge height measurements, the fraction of uncrosslinked chains and molecular weight distribution through Gel Permeation Chromatography (GPC), and material viscoelasticity through precise measurements of droplet sliding speeds on the substrate. Our work puts forth PDMS based organogels as a potential facile pathway towards enhanced condensation on soft substrates.

## 2. Experimental details

### 2.1 Materials and Sample Preparation

A series of soft substrates with a range of stiffness and fraction of uncrosslinked chains were prepared by varying the PDMS (Sylgard 184, Dow Corning) base to curing agent (crosslinker) ratio and adding lubricant (silicone oil, Xiameter PMX 200, 100 cSt, Credimex Inc.) prior to curing. For each sample, the base, the crosslinking agent and the lubricant were first thoroughly mixed in the desired weight ratio and then degassed for 15 to 20 minutes in a vacuum desiccator to remove air bubbles. Subsequently, the mixture was poured into a mold and then cured in an oven at 80°C for 2 hours. The thickness of all samples was 2 mm. In essence, our samples can be regarded as bulk PDMS. In order to ensure consistent curing conditions for all samples, we scheduled our experiments such that sample storage time between the end of curing and the start of experiments was maintained nearly the same across all the samples.

In the subsequent sections, each substrate is designated as 'PxLy' where 'x' is the base to crosslinker ratio i.e. $x = Ba/Cr$ and 'y' is the percentage by weight of lubricant defined as $y = 100Lu/(Lu + Cr(1 + x))$. Here, $Ba$, $Lu$ and $Cr$ are the weights of base, lubricant and crosslinker in the uncured mixture, respectively. For example, while P10L0 represents PDMS with a 10:1 base to crosslinker ratio and no additional silicone oil lubricant, P10L5 represents PDMS with the same base to crosslinker ratio but 5% lubricant added by weight.

### 2.2 Sample Characterization

We characterized the soft substrates with respect to various material properties as described in the following.

**a) Fraction of uncrosslinked chains and molecular weight distribution**

Even after curing, PDMS samples contain uncrosslinked chains (29–31). To determine the relative amount and the molecular weight distribution of the uncrosslinked chains we dissolved out the oligomers. Therefore, each sample was kept immersed in a beaker with ~ 450-600 ml toluene, which was stirred at 200 rpm (round per minute) for ten days. During this period, uncrosslinked chains diffused into toluene, which is a good solvent for PMDS, the curing agent and silicone oil. The toluene was renewed at the end of the 5[th] and the 7[th] day of immersion. Subsequently, the sample was immersed in hexane for seven days, wherein the hexane was renewed at the end of the 2nd and 4th day. Hexane is more volatile than toluene which helps in the eventual drying of the sample. The sample was dried under a fume hood for one night, in the oven at 50°C, and eventually under vacuum for about 7 to 8 hours. Finally, the sample was cooled at room temperature and weighed again to calculate the fraction of uncrosslinked chains. Additionally, we performed GPC analysis in order to determine the composition of uncrosslinked chains in various PDMS substrates. For each sample, Gel Permeation Chromatography (GPC) analysis was performed using the toluene from first extraction which yielded the molecular weight distribution of the uncrosslinked chains extracted from that sample. (refer to Supplementary Information, Section 1 for details on GPC analysis).

**b) Substrate elasticity**

Water droplets condensed on soft substrates pull the substrate at the contact line while the substrate is depressed in the contact area between the droplet and the substrate due to Laplace pressure. The three interfacial tensions, $\gamma_{sl}$, $\gamma_{sv}$, and $\gamma_{lv}$ balance in a Neumann's triangle configuration at the contact line and a wetting ridge is formed around the droplet (32, 33). This wetting ridge influences the droplet coalescence and shedding process during condensation on soft substrates (14). Here, $\gamma_{sl}$, $\gamma_{sv}$, and $\gamma_{lv}$ represent surface tensions for solid-liquid, solid-vapor and liquid-vapor interfaces. Since the wetting ridge formation is dependent on substrate elasticity, we characterized the elasticity of each sample by using a micro-indentation test. This test enabled the measurement of the Young's modulus ($E$) of the substrate and was performed using the micromechanical testing station FT-MTA02 (FemtoTools AG, Buchs ZH, Switzerland). The substrate shear modulus $G$ is related to elastic modulus as $G = E/2(1 + \nu)$, where $\nu$ is the Poisson's ratio of the material. For elastomers, this relationship can be simplified to $E = 3G$ (17). (refer to Supplementary Information Section 2 for further details).

**c) Wetting ridge height**

We also independently characterized the wetting ridge formation by measuring the wetting ridge height using white light interferometer (Zygo Inc.)(34). For these measurements, droplets of ethylene glycol (Sigma Aldrich) dyed with 2.6% by weight Sudan IV (Sigma Aldrich) were used (35). The higher boiling point of ethylene glycol compared to water minimizes droplet evaporation at room temperature during the wetting ridge height measurements. Since the measured wetting ridge height is a function of droplet size (36), measurements for each substrate were repeated with droplets of multiple sizes ranging from ~4 µL to ~20 µL to ensure consistent comparison between substrates (see Supplementary Information figure S3).

**d) Viscoelastic dissipation in the substrate**

We characterized the viscoelastic dissipation of each substrate by letting small droplets of a range of sizes to slide on a soft substrate (18, 35). As the droplets slide, the soft material continuously deforms and relaxes in the wetting ridge traveling with the droplet. Analysis of the droplet sliding velocity as a function of droplet volume provides a relative comparison of substrate viscoelasticity in the wetting ridge. In our experiments, 0.5 to 3.5 µL droplets of ethylene glycol were allowed to slide on vertically oriented substrates while their motion was carefully monitored using a CMOS camera. For each droplet size, the droplet was deposited on the vertically oriented substrate. The droplet started sliding downwards and the droplet velocity was measured when it attained a constant (terminal) sliding velocity. For repeatability, two samples of each type were fabricated and tested under the same conditions. Further details are provided in Supplementary Information Section 4.

**2.3 Microscale condensation**

To characterize droplet nucleation density and individual droplet growth rates we optically observed *in-situ* microscale condensation on various samples. Therefore, the sample was enclosed in a custom-designed chamber for observations under an optical microscope (refer to Supplementary Information Figure S5 for a schematic of the experimental setup). Inside the chamber, an environment with 100% relative humidity was maintained by placing water-saturated paper wipes. The sample temperature was controlled by mounting it on a temperature-controlled microscope stage (BCS-196, Linkam Scientific) using a copper stub and thermal paste. The humidity and temperature of the environment inside the chamber were continuously monitored using a humidity sensor (HYT221, IST Inc.) and a temperature sensor (PT10000, IST Inc.). The condensation process was recorded with a CMOS camera, and the recorded videos were analyzed to determine nucleation density and droplet growth rate. Once the sample was in position, the chamber was sealed, and the temperature-controlled stage was set to 35°C in order to induce evaporation of water from wet paper wipes. Once a stable humidity level of 100% was achieved, the temperature setpoint of the stage was reduced to 20 °C to initiate condensation. Image acquisition was initiated as soon as the smallest droplet resolvable by the microscope appeared on the substrate. The imaging was continued at a constant acquisition rate until droplets started coalescing. It was ensured that all samples were tested under a consistent humid air chamber atmospheric temperature of 28.7 ± 0.5 °C. For each sample, at least three independent measurements of nucleation density and droplet growth rate were obtained.

**2.4 Dewing**

Substrates were tested for water collection through dewing by using a custom-designed setup. In a dewing experiment, PDMS substrates with and without bulk lubricant infusion were tested together so as to achieve consistent temperature and humidity conditions across the samples. The tested substrates were fabricated by crosslinking directly on a copper plate to ensure good thermal contact between PDMS samples and copper. Each sample was fabricated to be 2 mm thick. The copper plate with the samples was then mounted onto a Peltier cooler and subsequently the samples were exposed to humid air in a closed chamber. The humidity inside the chamber was maintained at ~70% by circulating humid air generated by passing compressed air through a bubbler. The Peltier cooler was set such that the copper plate temperature was maintained as 2°C. The resulting condensation on the samples was monitored by using a DSLR camera, and water condensed on the substrates was collected. Each experiment was run for ~ 17 hours, and copper plate temperature, chamber temperature, and humidity were continuously monitored. Subsequently, the collected water from each sample was weighed. (Refer Supplementary Information Section 6 for further details).

**2.5 Cloaking**

It has been observed that low surface tension lubricants can cloak the water droplets (22, 37). We used a laser scanning confocal fluorescent microscope (Leica TCS SP8) to directly visualize the cloaking layer covering drops and confirm this effect for PDMS samples with and without additional bulk lubricant infusion. Fluorescent dye (excitation maximum $\lambda_{ex} = 663$ nm, emission maximum $\lambda_{em} = 712$ nm) was added to the PDMS mixture during sample preparation. The mixture was spin-coated for 1 minute at 1100 rounds per minute onto 170 μm thick microscope coverslips. PDMS films were allowed to crosslink at 80°C for two hours. Afterwards, a 0.3 μL drop of 57% water and 43% glycerol (fluorescently labelled with Atto 488: $\lambda_{ex} = 504$ nm and $\lambda_{em} = 521$ nm) was placed and left for 30 minutes to ensure enough time for the formation of any cloaking layer. The addition of glycerol reduced evaporation and improved refractive index matching with PDMS, thus reducing undesired optical artefacts ($n_{water} = 1.33$, $n_{glycerol} = 1.47$, $n_{drop} = 1.41$, $n_{PDMS} = 1.41$, $n$: refractive index). Subsequently, the top part of the drop was visualized with a 20 × / 0.75 glycerol immersion objective using an argon 488 nm laser to excite the dye in the drop and a helium-neon 633 nm laser to excite the dye in the PDMS. The emitted fluorescent signals were simultaneously captured by separate detectors, and a final image showing the signal from each fluorescent dye was constructed.

## 3. Results & Discussion

**3.1 Substrate elasto-capillarity and composition:** Figure 1(a) shows the range of substrate elasticities and the corresponding wetting ridge heights explored. The wetting ridge height increases for softer substrates as it scales as $\sim \gamma \sin(\theta)/G$, where $\gamma$ is the liquid-vapour interfacial tension, and $\theta$ is the apparent contact angle with respect to the undeformed surface of the substrate. The wetting ridge height increases from ~ 100 nm to ~ 1.4 µm as the shear modulus is reduced from ~ 500 kPa to ~ 15 kPa (Fig. 1a). This is in line with the previously reported trends about large wetting ridge heights formed on soft surfaces (14). It is evident that the substrate with stoichiometric base to crosslinking ratio (10:1) is stiffer than the ones with non-stoichiometric ratios (see P10L0 vs. P1L0 and P40L0). Additionally, irrespective of the base to crosslinker ratio, the addition of lubricant to the bulk results in reduced stiffness of the cured polymer (for example, compare P10L0 to P10L5 and P10L25). Thus, deviation from stoichiometric mixing ratios as well as bulk lubricant infusion reduces the shear modulus of the substrate. Figure 1(a) also shows the percentage of uncrosslinked chains as extracted by swelling in toluene. We observe that for a given ratio of base to crosslinker, the percentage of uncrosslinked chains increases with bulk lubricant infusion (for instance, compare P10L0 with P10L5, and P10L25 and P1L0 with P1L5 and P1L25). Additionally, for any percentage of the lubricant, deviation from stoichiometric base to crosslinker ratio results in a higher percentage of uncrosslinked chains (for instance, compare P10L0 with P1L0 and P40L0, and P10L5 with P1L5).

Next, we characterize the composition of uncrosslinked chains in various samples through GPC analysis as, apart from the fraction of uncrosslinked chains, the composition of such chains is also likely to affect the viscoelastic dissipation from the substrate. Figure 1(b) shows results of GPC analysis of the three precursors used in the fabrication of samples – two components of the of Sylgard 184 kit i.e., base and crosslinker, and lubricant (Credimex Xiameter Silicone oil). The molecular weight distribution of the three components is significantly different. The lubricant Credimex has a unimodal molecular weight distribution centered around $10^4$ g/mol. Both the base and the crosslinker have a bimodal molecular weight distribution. The main peak of the crosslinker is centered around 1.6 x $10^4$ g/mol. However, the crosslinker also contains a small fraction of high molecular weight components, which are centered around 6 x $10^4$ g/mol. The molecular weight distribution of the base is centered around 5 x $10^4$ g/mol. However, the base also contains a small fraction of molecules having a molecular weight as low as $10^3$ g/mol.

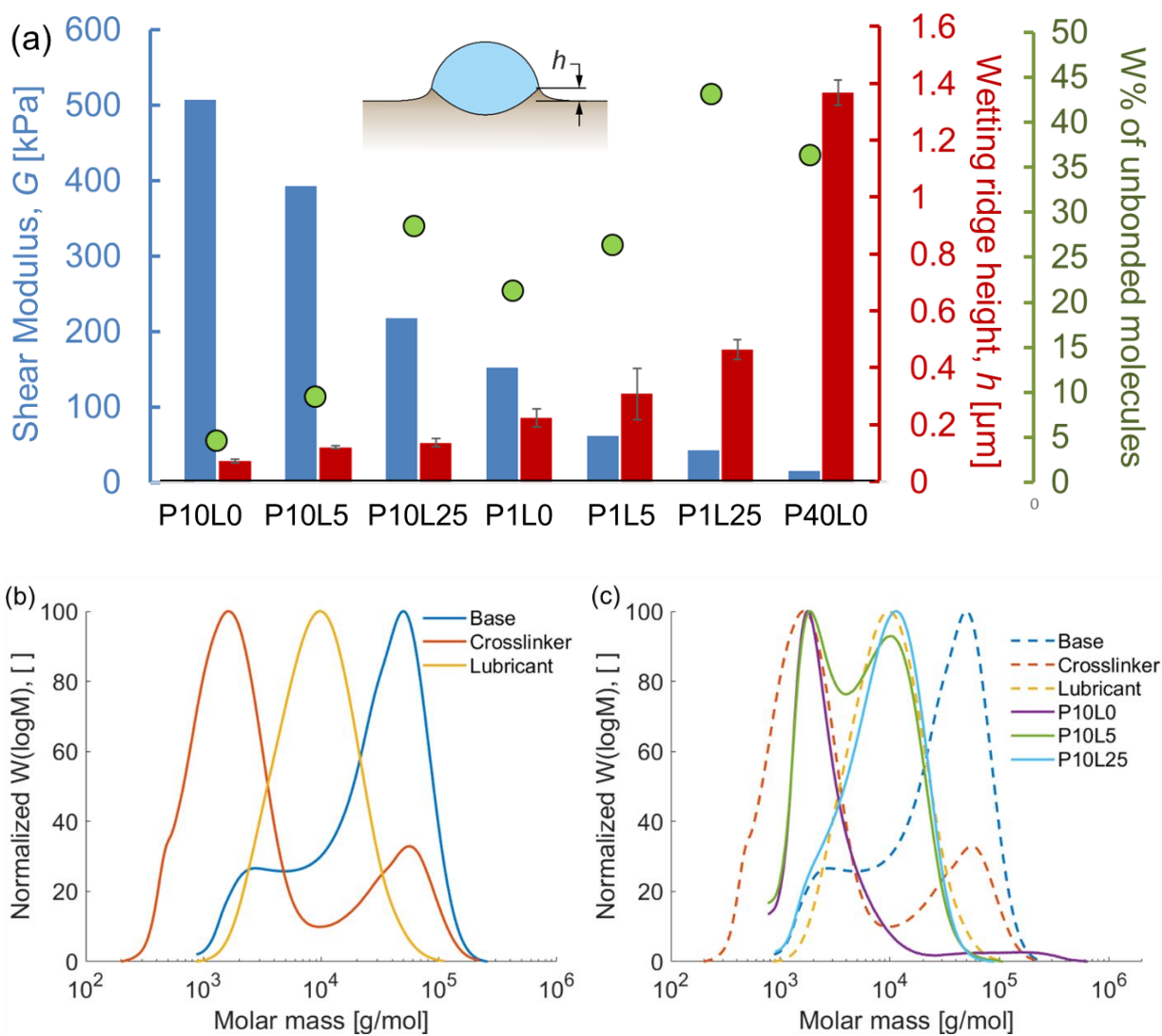

**Figure 1: (a)** Shear modulus (blue bars), wetting ridge heights (red bars) and percentage by weight of uncrosslinked chains (green dots) for the soft substrates. Inset figure defines the wetting ridge height ($h$), as measured by white light interferometry. For all substrates, wetting ridge height is measured using droplets of volume volume ~12μL or larger. Each substrate is designated as 'PxLy' where 'x' is the base to crosslinker ratio and 'y' is the percentage by weight of lubricant in the substrate. Substrates with lower shear modulus result in larger wetting ridge height. Deviation from stoichiometric base-to-crosslinker ratio and addition of lubricant results in a larger fraction of uncrosslinked chains. **(b)** GPC of precursors showing molecular weight distribution of uncrosslinked chains in base, crosslinker, and lubricant **(c)** Molecular weight distribution of uncrosslinked chains in P10L0, P10L5, and P10L25 (solid curves) overlaid on the distributions for precursors shown in (b) (dashed curves).

Figure 1(c) shows the molecular weight distributions of uncrosslinked chains for the substrates P10L0, P10L5 and P10L25 overlaid on the distributions shown in Fig. 1(b). The uncrosslinked chains diffuse from the substrate into toluene during swelling of the PDMS network in toluene. P10L0 does not contain Credimex. Since the peak is centered around 1.7 x $10^3$ g/mol, the uncrosslinked chains in P10L0 are due to the low molecular weight components of the base and of crosslinker. Both, the uncrosslinked chains in P10L5 and P10L25 show a peak centered around 6 x $10^4$ g/mol, which can be assigned to the added lubricant. In case of P10L5, the second peak centered around 1.7 x $10^3$ g/mol corresponds to the peak observed for the position of the low molecular weight peak of the crosslinker. However, the low molecular weight components of the Sylgard Base may also contribute to this peak. The uncrosslinked molecules in P10L25 mainly result from the lubricant. Similarly, the uncrosslinked chains in other samples are also governed by their respective compositions (refer Supplementary Information Section 1 for results of GPC analysis for P1L0, P1L5, P1L25 and P40L0). In essence, the above analysis indicates that the mobile uncrosslinked chains in PDMS substrates are dominated by the low molecular weight components of crosslinker and base and by the additional lubricant.

**3.2 Substrate viscoelasticity:** Droplets sliding down a vertical soft substrate attain a terminal velocity when the viscoelastic dissipation balances the work done by the driving force of droplet weight (17). Measuring this velocity for droplets of various volumes provides an estimation of the inherent viscoelastic dissipation in the substrate. Figure 2(a) compares droplet velocity $(U_{drop})$ as a function of droplet volume $(V_{drop})$ for P10L0, P1L0, and P40L0. Sliding droplets of same size attain higher sliding velocities on P1L0 than P10L0 and lower sliding velocities on P40L0 than P10L0. Figure 2(b) shows the same comparison for P10L0, P10L5 and P10L25. Here, sliding droplets on substrates with bulk lubricant infusion, P10L5 and P10L25, attain much higher velocities than P10L0. Moreover, this increasing in sliding velocity over P10L0 is much higher than that obtained in the case of P1L0 in Fig. 2(a). This clearly indicates that bulk lubricant infusion results in a significant reduction in viscoelastic dissipation in soft substrates. A comparison on Fig. 2 and Fig.1 leads to three conclusions. First, viscoelastic dissipation can be reduced by infusing the PDMS with lubricant despite the fact than lubricant infused substrates have lower shear modulus (and thus larger wetting ridge formation) than P10L0. Second, while viscoelastic dissipation can be reduced by increasing the proportion of crosslinker (see P1L0 vs P10L0 in Fig 2a), this reduction is much less significant than obtained by infusing lubricant in PDMS P10L0. And third, as eluded to in section 3.1, viscoelastic dissipation depends not only on the fraction of uncrosslinked chains but also on the composition of the chains. For instance, P40L0 has a much higher fraction of uncrosslinked chains than P10L0, P10L5 and P10L25 (Fig. 1a), yet viscoelastic dissipation is higher in P40L0. As shown by GPC analysis in Fig 1c, lubricant infused in P10L5 and P10L25 contributes to uncrosslinked chains and makes the substrate more slippery.

In each figure, dotted lines are curve fits based on an energy model. This model assumes that the work done by the weight of sliding drop is mainly dissipated through viscoelastic breaking by the substrate. Although droplets sliding on PDMS infused with lubricant may also experience viscous dissipation from a layer of lubricant formed on the top surface of the sample (38), it is difficult to distinguish between the viscous dissipation from lubricant and viscoelastic dissipation from the elastomer. In essence, the viscoelastic dissipation in our model represents the net energy dissipation caused by the substrate. The drop sliding velocity, $U_{drop}$ can be related to volume of drop, $V_{drop}$ as $U_{drop} \approx U_0 \left(\frac{\pi \rho V_{drop} g G h}{2 r \gamma^2 I}\right)^{\frac{1}{m}}$ (17). Next, in order to quantitatively estimate differences in viscoelastic dissipation among different substrates, we compare the ratio of viscoelastic dissipation in the substrate due to the movement of the contact line $(\dot{D}_{ve})$ to viscous dissipation $(\dot{D}_v)$ within the droplet. This dissipation ratio is given by $\frac{\dot{D}_{ve}}{\dot{D}_v} = \frac{\gamma^2 U_{drop}^{m-1} I \theta}{3\pi^2 G h \mu U_0^m l}$ (17, 34) and is lower for more slippery soft substrates. Here, $\rho$, $\gamma$ and $\mu$ are density, surface tension and viscosity of the liquid respectively, $\theta$ is the apparent contact angle, $U_0$ and $m$ are rheological constants for a substrate, $I$ and $l$ are constants based on droplet morphology, and $r$ is the base radius of the drop.

Based on the droplet sliding measurements shown in Fig 2b, we estimate that, relative to P10L0, $\frac{\dot{D}_{ve}}{\dot{D}_v}$ reduces by a factor of ~28 for P10L5 and ~57 for P10L25 for a 5µl drop, thus indicating a large reduction in viscoelastic dissipation for lubricant infused substrates. However, on P1L0 the addition of lubricant did not result in any significant change in viscoelastic dissipation, unlike the case of P10L0, indicating that bulk lubricant infusion in stoichiometric base-to-crosslinker provides the most benefit in terms of achieving slippery soft surfaces. In other words, lubricant addition facilitates sliding on sufficiently cross-linked PDMS (i.e. PDMS with stoichiometric base to crosslinker ratio). This is likely due to the much larger relative contribution of crosslinker to uncrosslinked chains in P1L5 and P1L25 compared to P10L5 and P10L25 (Figure 1(c) and Figure S1). Refer to Supplementary Information Section 6 for details of the model fit, estimation of the dissipation ratio and comparison of droplet sliding velocities for P1L0, P1L5 and P1L25.

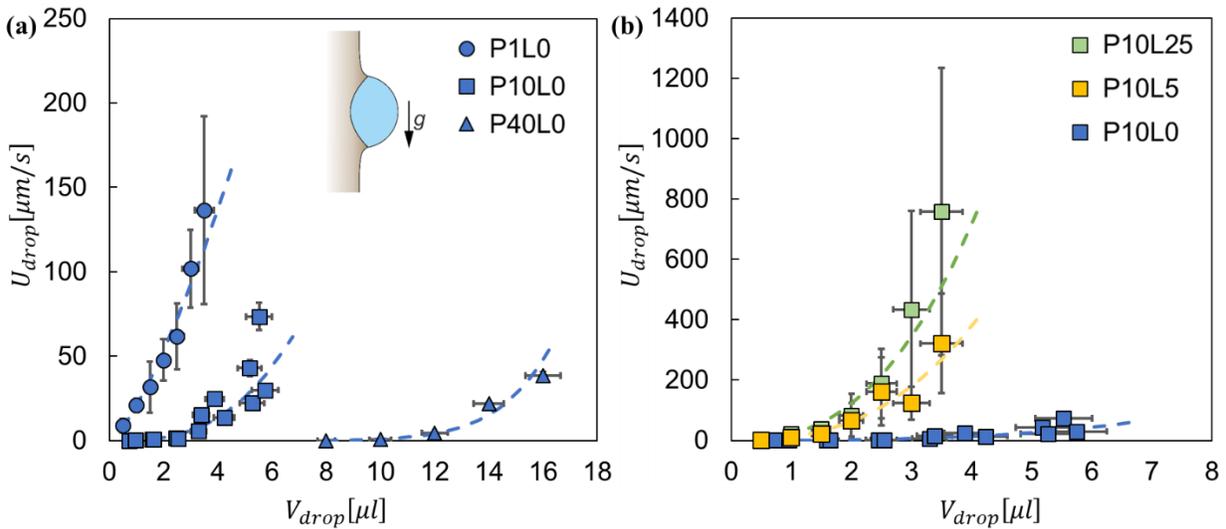

**Figure 2** (a) Comparison of droplet sliding velocities as a function of droplet volume for P10L0, P1L0 and P40L0. Inset figure shows a schematic of drop sliding on the substrate under gravity. (b) Similar comparison for P10L0, P10L5 and P10L25. Higher sliding velocities for a given droplet volume indicate lower viscoelastic braking in the substrate. Dotted lines represent curve fits with $U_{drop} = \frac{U_0}{V_0^{2/3m}} V_{drop}^{2/3m}$ where $V_0$, $U_0$ and $m$ are fitting parameters. Values of these parameters are listed in Table S1.

**3.3 Microscale condensation:** Microscale *in-situ* condensation observations yielded quantitative information on nucleation and subsequent growth of condensate droplets on various substrates. Figure 3 shows nucleation density ($N$) of heterogeneous condensation as a function of the substrate shear modulus. Overall, the nucleation density follows an inverse relation with substrate shear modulus (14, 39) wherein $N$ changes more steeply at lower shear modulus values. Additionally, substrates with bulk lubricant infusion achieve higher nucleation density. This is evident from comparison of $N$ for P10L0 with P10L5 and P10L25, and similarly from comparison of P1L0 against P1L5 and P1L25. Interestingly however, we find that for substrates with bulk lubricant infusion, $N$ values show minimal dependence on the percentage of lubricant infused in the material. For instance, P10L5 and P10L25 achieve nearly similar nucleation density despite the fact that the shear modulus for P10L25 is nearly 50 percent lower and fraction of uncrosslinked chains is nearly 200 percent higher (inset plot in Fig.3) than that for P10L5. Based on this insensitivity of $N$ value to change in bulk lubricant concentration, we speculate that the lubricant-dominated uncrosslinked chains in these substrates (see Figure 1(c) and Figure S1) diffuse outwards from the bulk and induce the formation of a continuous layer of lubricant on the top surface of the substrate. This lubricant layer in turn governs the nucleation dynamics during condensation, irrespective of the amount of lubricant present in the bulk of the material.

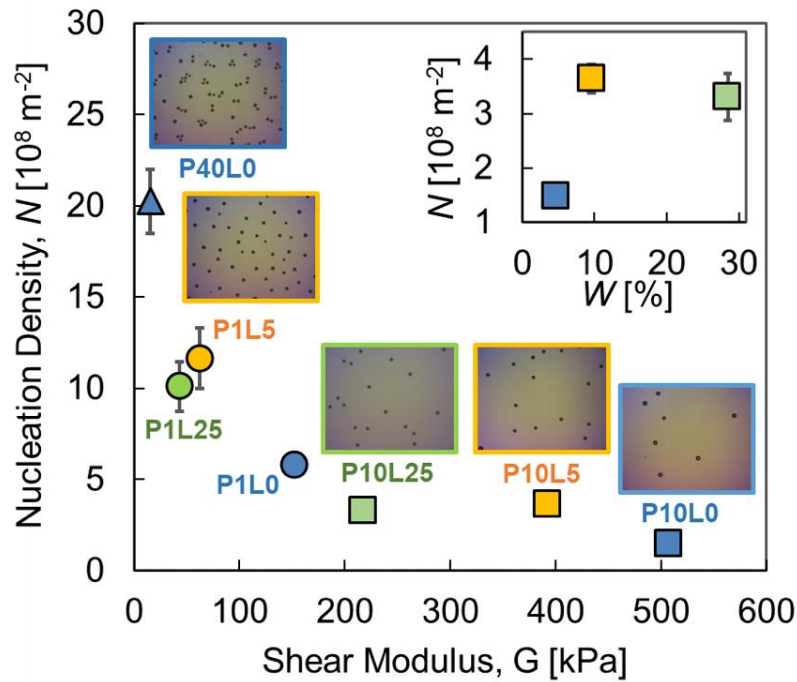

**Figure 3:** Nucleation density of condensation on various substrates at 100% humidity and room temperature as a function of substrate shear modulus. Squares correspond to substrates with 10:1 base to crosslinker ratio (i.e. P10L0, P10L5 and P10L25), circles correspond to substrates with 1:1 base to crossslinker ratio (i.e. P1L0, P1L5 and P1L25). Inset figures show typical condensation

nucleation densities on four substrates over an observation area of 235 µm x 191 µm. Inset plot shows nucleation density as a function of percentage of uncrosslinked chains for P10L0, P10L5 and P10L25.

Figure 4(a) shows pre-coalescence growth of condensate droplets for P10L0, P10L5 and P10L25 as a function of observation time. Each curve is an average of growth curves of three different randomly chosen droplets for the respective substrates. The observation is initiated when first droplets become visible at the limit of resolution of the optical system. The observation is continued while the temperature of the cooling stage is held constant at 20 ℃ and the humidity and temperature in the experimental chamber are maintained at 100% and 28.7 ℃ respectively. Interestingly, the droplet growth rate for non-lubricated P10L0 is higher than that for the lubricated samples P10L5 and P10L25. Thus, while the lubricated substrates show higher performance than P10L0 in terms of nucleation density, the opposite is true in terms of droplet growth. Figure 4(b) presents snapshots of the condensate droplet growth process on P10L0 and P10L5 demonstrating the slower droplet growth for P10L5. Moreover, the droplets grow at nearly the same rate on P10L5 and P10L25 even though P10L25 has nearly 5 five times more lubricant, and thus shows increased substrate deformability, than P10L5. The comparison of pre-coalescence growth rate between P1L0, P1L5 and P1L25 is similar (refer Supplementary figure S7). This similarity of droplet growth, along with the earlier stated similarity in nucleation density for these widely different lubricant concentrations, again indicates that the condensation dynamics for lubricated substrates are governed by a surface characteristic that dominates the large difference in bulk material properties. As already alluded to, we believe that this surface property is the existence of a distinct layer of lubricant formed on the surface by the outwards diffusion of the lubricant from the bulk elastomer matrix. The presence of this lubricant layer controls the nucleation density as well as the condensate droplet growth rate.

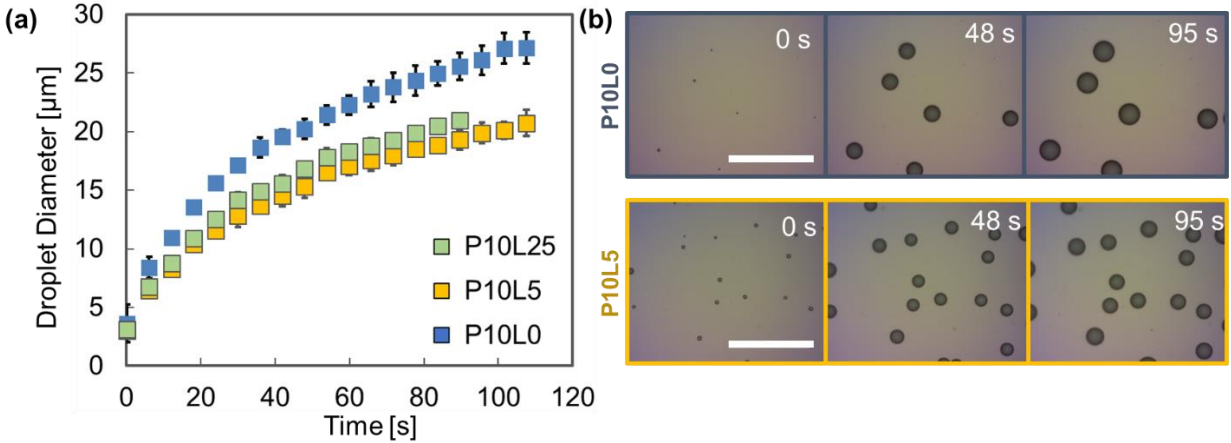

**Figure 4:** (a) Comparison of pre-coalescence condensate droplet growth rate on substrates without (P10L0) and with (P10L5 and P10L25) bulk lubricant infusion (b) Snapshots of condensation on P10L0 and P10L5 at equivalent time-instants after the start of nucleation, under same atmospheric humidity and surface subcooling conditions. Scale bars in white represent 100 µm.

**3.4 Detection of cloaking layer and cloaking kinetics** The reduced condensate droplet growth rate for substrates with bulk lubricant infusion points towards the formation of a cloaking layer over the droplets, similar to the case of lubricant-infused rigid textures (22). For the lubricant used i.e. Xiameter PMX 200/100cs Silicone oil, cloaking of water droplets is energetically favourable as, it has a positive spreading coefficient on water ($S = \gamma_{wv} - \gamma_{lv} - \gamma_{lw} > 0$), where $\gamma_{lv}$, $\gamma_{lw}$ and $\gamma_{wv}$ are the interfacial tensions for lubricant-air, lubricant-water and water-air interfaces (40). Thus an oil cloaking layer with thickness ~ 100 nm would be formed on the condensing drops (22). It has been speculated that a cloak may form a barrier for diffusion of water molecules from vapour to liquid thus slowing the overall process of phase change (21, 22).

To confirm the formation of a cloaking layer on water droplets, we imaged 0.3 µL sized drops placed on P10L0 and P10L25 using confocal microscopy (Figure 5, taken around 30 minutes after placing the drop). A side view cross section of the cloak (PDMS, orange) is shown in Figure 5(a), with the relevant region highlighted in the schematic drawn in Figure 5 (b). Figures 5 (c)-(e) show the top part of the drop. Fluorescent signal from the glycerol/water drop is shown in blue [Figure 5 (c)] and fluorescence from the untethered PDMS chains cloaking the drop is shown in orange [Figure 5 (d)]. These two signals are overlaid into a single image in Figure 5 (e). While the drops were initially bare without any observable lubricant cloak, after around 30 minutes, they became cloaked by material from the substrates for both P10L0 and P10L25. This shows that irrespective of the concentration of the additional lubricant, the drops are ultimately cloaked. While for P10L0 this cloaking proceeds by the diffusion of the uncrosslinked Slygard 184 chains from the substrate, for P10L25, it is a combination of both the diffusion of the uncrosslinked Sylgard 184 chains and the diffusion of the added lubricant (Xiameter).

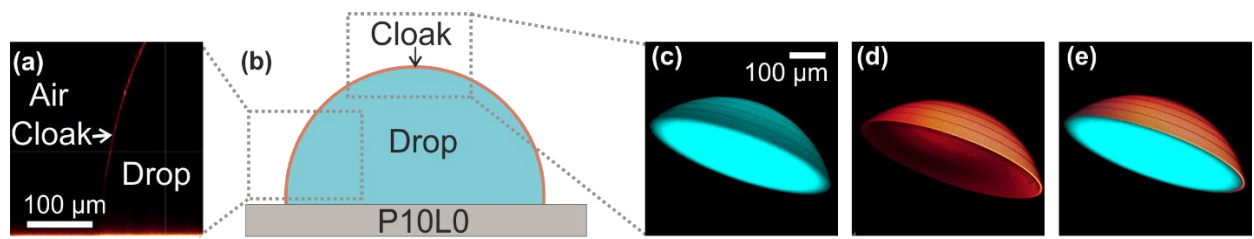

**Figure 5 Visualization of the PDMS cloak on a drop using laser scanning confocal microscopy.** (a) 2D cross section of the cloak (Orange, custom made perylene fluorescent dye ) around a 57% glycerol- 43% water by weight drop. (b) Schematic of cloaked drop on P10L0. (c), (d) and (e) 3D confocal image of the top of a glycerol-water drop (cyan, fluorescent dye: Atto 488) taken approximately 30 minutes after the drop was placed on the surface. (c) shows fluorescent signal from the drop, (d) shows fluorescent signal from the cloak, and (e) is an overlay of (c) and (d). The images were taken with a $20 \times/0.75$ glycerol immersion objective with horizontal resolution of ≈0.3 µm and vertical resolution of ≈1 µm. Similar result is obtained for visualization of cloaked drop on P10L25.

While confocal microscopy confirms the formation of cloaking layer on water droplets, it does not provide reliable quantitative information on how fast the cloaking proceeds on various substrates. Cloaking is expected to reduce the surface tension of the drop. Therefore, to quantify differences in the cloaking kinetics due to lubricant infusion in PDMS, we measured the change in drop-air interfacial tension of a water drop placed into contact with P10L0, P10L5 and P10L25 as follows (41). A glass slide with a 2 mm diameter hole was spin-coated with the PDMS mixture [Fig. 6 (a)]. A 38 µL drop of pure water was positioned in pendant configuration such that its three-phase contact line was pinned at the edge of the hole. This is to ensure that the contact line does not slide on the substrate as the interfacial tension of the drop changes (refer Supplementary Information Section 8 for further experimental details). Initially, the drop's interface consists of water molecules surrounded by vapour (air), resulting in a surface tension equal to that of pure water ($\gamma_{wv}$). However, due to the positive spreading coefficient of PDMS on water (40), the drop draws PDMS from the substrate and is cloaked. This results in a decrease in the effective surface tension of the drop-vapour interface from $\gamma_{wv}$ to $\approx \gamma_{lv} + \gamma_{lw}$. To obtain the interfacial tension of the drop as a function of time [Fig. 6 (b)], we imaged the drop's shape and fitted the Young-Laplace equation to its contour [Fig. 6 (c)]. As shown in Figure 6 (b), the surface tension remained unchanged for a certain time, then declined and eventually reached a plateau. This transition can be interpreted as corresponding to the formation of a continuous layer of the lubricant over the droplet surface (22, 42, 43). For P10L25, the transition occurred within 30 s whereas for P10L5 it took several minutes. No change in interfacial tension was observed on P10L0 over 30 minutes. These results clearly show a significant difference in kinetics of cloaking between substrates having different amounts of mobile PDMS molecules arising from the uncrosslinked chains, either due to the bulk lubricant infusion or due to uncrosslinked chains from the Sylgard 184 mixture. Hence, the growing condensate droplets on substrates with bulk lubricant infusion would stay uncloaked for much lesser time thus reducing the overall droplet growth rate compared to substrates without bulk infusion.

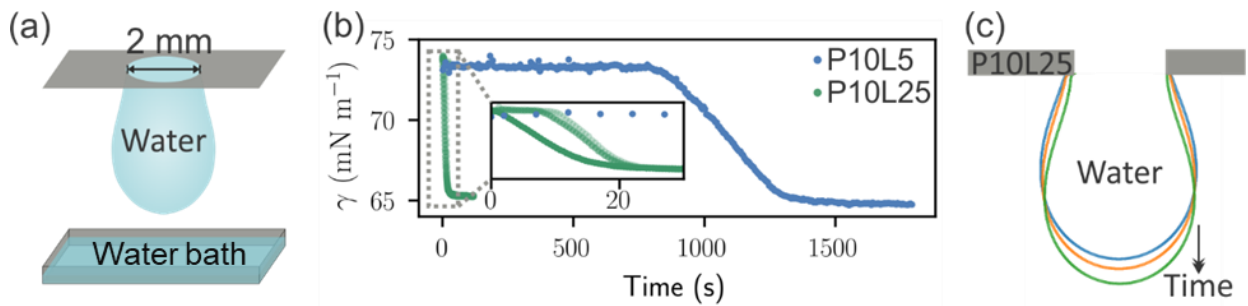

**Figure 6 Measuring cloaking timescale using pendant drop method.** (a) Schematic of setup. A 2 mm diameter hole was drilled into a glass slide coated with P10L0, P10L5 or P10L25. A water drop (38 μL) was suspended such that the drop's contact line was pinned at the hole's circumference. A water bath was placed a few millimeters below the drop to suppress evaporation. The drop profile was monitored over time, and its surface tension calculated and plotted in (b). The inset in (b) focusses on the transition for P10L25. In the inset, the different green lines correspond to repeats of the same experiment. (c) Evolution of the drop contour over time on P10L25. No change in surface tension was observed on P10L0 over 30 minutes (see Supplementary Figure S8).

**3.5 Dewing:** We evaluated three different substrates, namely P10L0, P10L5 and P10L25 to capture the effect of lubrication on water collection rates. P10L25 and P10L5 showed similar average dew water collection rates over the time period of experiment (17 hours) and achieve about 44% and 49% higher water collection respectively than P10L0 under similar dewing conditions (Fig. 7, blue bars). The higher water collection on P10L5 and P10L25 arises from a shorter delay time for the initiation of condensate droplet shedding compared to P10L0 (Fig. 7, orange points). Additionally, we calculate that after the onset of droplet shedding, average water collection rate on P10L5 and P10L25 is around 35% higher compared to P10L0. Supplementary video shows typical dewing on P10L0, P10L5 and P10L25 over the course of one experiment.

The enhanced phase change on PDMS substrates with bulk lubricant infusion show that the higher nucleation density and lower viscoelastic dissipation more than compensate for lower individual droplet growth rate. As a result, it is evident that bulk lubricant infusion is a promising approach to enhance water condensation on soft substrates by reducing viscoelastic braking while retaining the inherent high nucleation density. We recognize that the collected water would contain a small volume fraction of uncrosslinked chains from PDMS and lubricant due to cloaking of water droplets (29) which may be removable to ensure water potability if so desired.

## 4 Conclusions

We demonstrate that water condensation on soft substrates can be significantly enhanced through bulk lubricant infusion in PDMS. Lubricant infusion in PDMS reduces the shear modulus and increases the fraction of uncrosslinked chains in the substrate. GPC analysis shows that the mobile chains in such substrates are dominated by the low molecular weight components of the crosslinker and base and by the additional lubricant. Additionally, a mere 5% by weight lubricant infusion reduces the viscoelastic dissipation in the substrate by nearly 28 times compared to crosslinked PDMS with stoichiometric composition. Moreover, lubricant infusion in stoichiometric PDMS composition provides the most benefit in terms of achieving slippery soft substrates compared to lubricant infusion in non-stoichiometric PDMS composition. Lubricant infused PDMS also achieves a higher density of droplet nucleation during condensation. However, individual condensate droplet growth rate through direct condensation is reduced on such substrates due to cloaking of the condensate droplets with the lubricant. The nucleation density and droplet growth rate are observed to be rather insensitive to the amount of lubricant in the substrate. Overall, slippery soft substrates studied, obtained by bulk lubricant infusion in PDMS, achieved up to 49% higher rate of water condensation compared to PDMS without any additional lubricant. The overall effect of bulk lubricant infusion in terms of enhancing the rate of water condensation provides an important pathway for realistic application of soft substrates in water condensation applications.

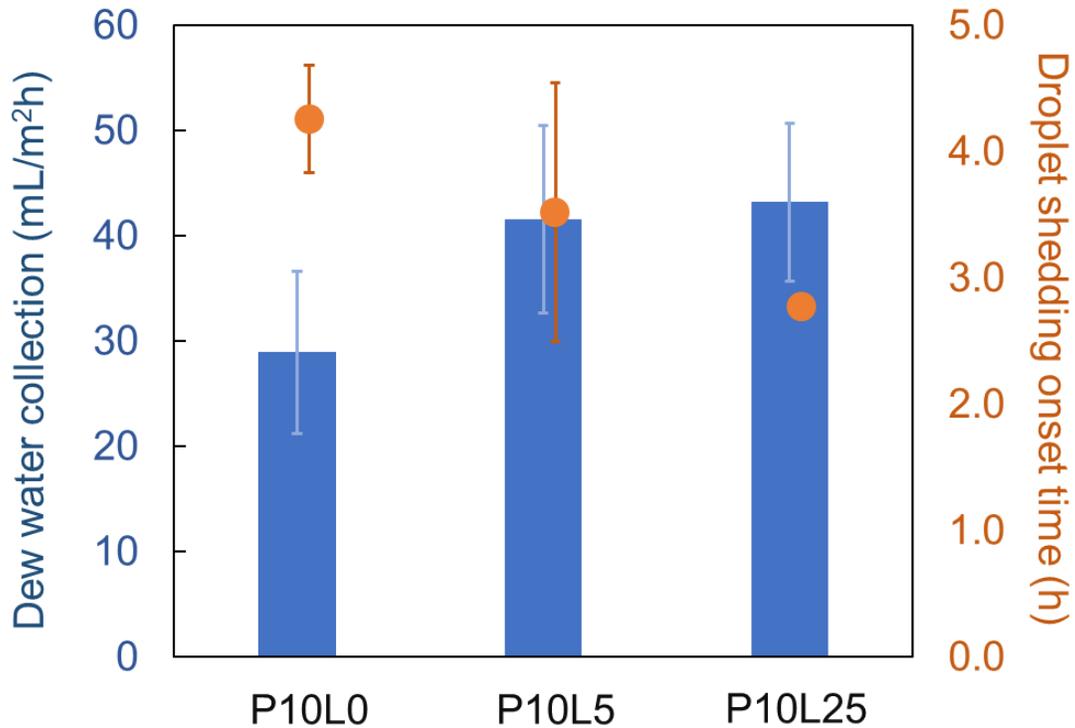

**Figure 7:** Dew water collection rate (blue bars) and onset time for droplet shedding (orange points) on P10L0, P10L5 and P10L25 under an atmospheric relative humidity, temperature of humid air and surface temperature ~70% and ~ 17℃ and copper plate surface temperature of ~2℃. Each bar and dot represents an average over four experiments with each experiment conducted over a period of 17 hours. Error bars represent standard deviation over these experiments. Dew water collection rate is average rate over the period of experiment. Droplet shedding time represents the time from start of experiment when first condensate drop sheds from the substrate.

**Conflicts of interest**

The authors declare no conflict of interest.

**Acknowledgements**

We thank Mr. Peter Feusi and Mr. Jovo Vidic for their help with the experimental setup assembly as well as Dr. Hadi Eghlidi for inputs on optical observations. This project has received funding from the European Union's Horizon 2020 research and innovation program under grant number 801229 (HARMoNIC), and through the innovative training network LubISS under the grant number 722497 (A.N., M.D'A., D.V.) and the Commission for Technology and Innovation, (CTI) under the Swiss Competence Centers for Energy Research (SCCER) program (Grant No. KTI.2014.0148).

**References**

1. Beér JM (2007) High efficiency electric power generation: The environmental role. *Prog Energy Combust Sci* 33(2):107–134.

2. González D, Amigo J, Suárez F (2017) Membrane distillation: Perspectives for sustainable and improved desalination. *Renew Sustain Energy Rev* 80:238–259.

3. Bintein P-B, Lhuissier H, Mongruel A, Royon L, Beysens D (2019) Grooves Accelerate Dew Shedding. *Phys Rev Lett* 122(9):98005.

4. Sharma CS, Lam CWE, Milionis A, Eghlidi H, Poulikakos D (2019) Self-Sustained Cascading Coalescence in Surface Condensation. *ACS Appl Mater Interfaces* 11(30):27435–27442.

5. Edalatpour M, Murphy KR, Mukherjee R, Boreyko JB (2020) Bridging-Droplet Thermal Diodes. *Adv Funct Mater* 30(43):2004451.

6. Rose JW (2002) Dropwise Condensation Theory and Experiment: A Review. *Proc Inst Mech Eng Part A J Power Energy* 216(2):115–128.

7. Enright R, Miljkovic N, Alvarado JL, Kim K, Rose JW (2014) Dropwise Condensation on Micro- and Nanostructured Surfaces. *Nanoscale Microscale Thermophys Eng* 18(3):223–250.

8. Sharma CS, Combe J, Giger M, Emmerich T, Poulikakos D (2017) Growth Rates and Spontaneous Navigation of Condensate Droplets Through Randomly Structured Textures. *ACS Nano* 11(2):1673–1682.

9. Boreyko JB, Chen C-H (2009) Self-Propelled Dropwise Condensate on Superhydrophobic Surfaces. *Phys Rev Lett* 103(18):184501.

10. Donati M, et al. (2021) Sprayable Thin and Robust Carbon Nanofiber Composite Coating

for Extreme Jumping Dropwise Condensation Performance. *Adv Mater Interfaces* 8(1):2001176.

11. Sharma CS, Stamatopoulos C, Suter R, von Rohr PR, Poulikakos D (2018) Rationally 3D-Textured Copper Surfaces for Laplace Pressure Imbalance-Induced Enhancement in Dropwise Condensation. *ACS Appl Mater Interfaces* 10(34):29127–29135.

12. Seo D, Lee J, Lee C, Nam Y (2016) The effects of surface wettability on the fog and dew moisture harvesting performance on tubular surfaces. 6:24276.

13. Lee A, Moon M-W, Lim H, Kim W-D, Kim H-Y (2012) Water harvest via dewing. *Langmuir* 28(27):10183–10191.

14. Sokuler M, et al. (2010) The Softer the Better: Fast Condensation on Soft Surfaces. *Langmuir* 26(3):1544–1547.

15. Eslami F, Elliott JAW (2011) Thermodynamic Investigation of the Barrier for Heterogeneous Nucleation on a Fluid Surface in Comparison with a Rigid Surface. *J Phys Chem B* 115(36):10646–10653.

16. Jerison ER, Xu Y, Wilen LA, Dufresne ER (2011) Deformation of an Elastic Substrate by a Three-Phase Contact Line. *Phys Rev Lett* 106(18):186103.

17. Shanahan MER, Carré A (2002) Spreading and dynamics of liquid drops involving nanometric deformations on soft substrates. *Colloids Surfaces A Physicochem Eng Asp* 206(1–3):115–123.

18. Karpitschka S, et al. (2015) Droplets move over viscoelastic substrates by surfing a ridge. *Nat Commun* 6:7891.

19. Wong T-S, et al. (2011) Bioinspired self-repairing slippery surfaces with pressure-stable omniphobicity. *Nature* 477(7365):443–447.

20. Lafuma A, Quéré D (2011) Slippery pre-suffused surfaces. *EPL (Europhysics Lett* 96(5):56001.

21. Anand S, Paxson AT, Dhiman R, Smith JD, Varanasi KK (2012) Enhanced Condensation

on Lubricant-Impregnated Nanotextured Surfaces. *ACS Nano* 6(11):10122–10129.

22. Anand S, Rykaczewski K, Subramanyam SB, Beysens D, Varanasi KK (2015) How droplets nucleate and grow on liquids and liquid impregnated surfaces. *Soft Matter* 11(1):69–80.

23. Dai X, et al. (2018) Hydrophilic directional slippery rough surfaces for water harvesting. *Sci Adv* 4(3):eaaq0919.

24. Kim P, Kreder MJ, Alvarenga J, Aizenberg J (2013) Hierarchical or Not? Effect of the Length Scale and Hierarchy of the Surface Roughness on Omniphobicity of Lubricant-Infused Substrates. *Nano Lett* 13(4):1793–1799.

25. Zhang P, Zhao C, Zhao T, Liu M, Jiang L (2019) Recent Advances in Bioinspired Gel Surfaces with Superwettability and Special Adhesion. *Adv Sci* 6(18):1900996.

26. Golovin K, et al. (2016) Designing durable icephobic surfaces. *Sci Adv* 2(3):e1501496.

27. Zhu L, et al. (2013) Ice-phobic Coatings Based on Silicon-Oil-Infused Polydimethylsiloxane. *ACS Appl Mater Interfaces* 5(10):4053–4062.

28. Yeong YH, Milionis A, Loth E, Sokhey J (2018) Self-lubricating icephobic elastomer coating (SLIC) for ultralow ice adhesion with enhanced durability. *Cold Reg Sci Technol* 148:29–37.

29. Hourlier-Fargette A, Antkowiak A, Chateauminois A, Neukirch S (2017) Role of uncrosslinked chains in droplets dynamics on silicone elastomers. *Soft Matter* 13(19):3484–3491.

30. Pham JT, Schellenberger F, Kappl M, Butt H-J (2017) From elasticity to capillarity in soft materials indentation. *Phys Rev Mater* 1(1):15602.

31. Wong WSY, et al. (2020) Adaptive Wetting of Polydimethylsiloxane. *Langmuir*. doi:10.1021/acs.langmuir.0c00538.

32. de Gennes P-G, Brochard-Wyart F, Quéré D (2004) *Capillarity and Wetting Phenomena* doi:10.1007/978-0-387-21656-0.

33. Style RW, Dufresne ER (2012) Static wetting on deformable substrates, from liquids to soft solids. *Soft Matter* 8(27):7177–7184.

34. Carre A, Gastel J-C, Shanahan MER (1996) Viscoelastic effects in the spreading of liquids. *Nature* 379(6564):432–434.

35. Karpitschka S, et al. (2016) Liquid drops attract or repel by the inverted Cheerios effect. *Proc Natl Acad Sci* 113(27):7403–7407.

36. Style RW, et al. (2013) Universal Deformation of Soft Substrates Near a Contact Line and the Direct Measurement of Solid Surface Stresses. *Phys Rev Lett* 110(6):66103.

37. Schellenberger F, et al. (2015) Direct observation of drops on slippery lubricant-infused surfaces. *Soft Matter* 11(38):7617–7626.

38. Keiser A, Baumli P, Vollmer D, Quéré D (2020) Universality of friction laws on liquid-infused materials. *Phys Rev Fluids* 5(1):14005.

39. Phadnis A, Rykaczewski K (2017) Dropwise Condensation on Soft Hydrophobic Coatings. *Langmuir* 33(43):12095–12101.

40. Smith JD, et al. (2013) Droplet mobility on lubricant-impregnated surfaces. *Soft Matter* 9(6):1772–1780.

41. Naga A, et al. (2021) How a water drop removes a particle from a hydrophobic surface. *Soft Matter* 17(7):1746–1755.

42. Langmuir I (1917) THE CONSTITUTION AND FUNDAMENTAL PROPERTIES OF SOLIDS AND LIQUIDS. II. LIQUIDS.1. *J Am Chem Soc* 39(9):1848–1906.

43. Bergeron V, Langevin D (1996) Monolayer Spreading of Polydimethylsiloxane Oil on Surfactant Solutions. *Phys Rev Lett* 76(17):3152–3155.

# Supplementary Information

# Enhancing condensation on soft substrates through bulk lubricant infusion

Chander Shekhar Sharma[2], Athanasios Milionis[1], Abhinav Naga[4], Cheuk Wing Edmond Lam[1], Gabriel Rodriguez[1], Marco Francesco Del Ponte[1], Valentina Negri[1], Hopf Raoul[3], Maria D'Acunzi[4], Hans-Jürgen Butt[4], Doris Volmer[4], Dimos Poulikakos[1†]

[1]Laboratory of Thermodynamics in Emerging Technologies, Department of Mechanical and Process Engineering, ETH Zurich, 8092 Zurich, Switzerland

[2]Thermofluidics Research Lab, Department of Mechanical Engineering, Indian Institute of Technology Ropar, Rupnagar, 140001 Punjab, India

[3] Institute of Mechanical Systems, Department of Mechanical and Process Engineering, ETH Zurich, 8092 Zurich, Switzerland

[4]Max Planck Institute for Polymer Research, Ackermannweg 10, D-55128, Mainz, Germany

†E-mail: dpoulikakos@ethz.ch. Phone: +41 44 632 27 38. Fax: +41 44 632 11 76

**Contents**

1. Gas permeation chromatography (GPC)
2. Determination of substrate shear modulus using micro-indention tests
3. Additional data on wetting ridge height
4. Droplet sliding experiments
5. Microscale condensation setup
6. Dewing experiments
7. Energy model and dissipation ratio
8. Pre-coalescence droplet growth rates on P1L0, P1L5 and P1L25
9. Experimental details to measure cloaking timescale
10. Supplementary video

## 1. Gas permeation chromatography (GPC)

GPC experiments were performed using an Agilent Technologies 1260 instrument consisting of a pump, autosampler and column oven. As eluent toluene was used. A column set consisting of 3 columns: SDV $10^6$ Å, SDV $10^4$ Å and SDV 500Å (PSS Standards Service GmbH, Mainz, Germany), all of 300 x 8 mm and 10μm average particle size were used at a flow rate of 1.0 mL/min and a column temperature of 30 °C. The injection volume was 100 μL. Detection was accomplished with a RI detector (Agilent Technologies). In order to reproduce the measurement, each sample was injected twice.

Data acquisition and evaluation was performed using PSS WINGPC UniChrom (PSS Polymer Standards Service GmbH, Mainz, Germany). Calibration was carried out by using the universal calibration method with polystyrene standards provided by PSS Polymer Standards Service GmbH (Mainz, Germany) and the Mark-Houwink coefficients for PDMS in toluene.

Figures S1(a) below shows result of GPC analysis for P1L0, P1L5 and P1L25. Similarly, to the discussion in main text, the lubricant contributes towards additional uncrosslinked chains in P1L5 and P1L25. Figure S1(b) shows corresponding results for P40L0.

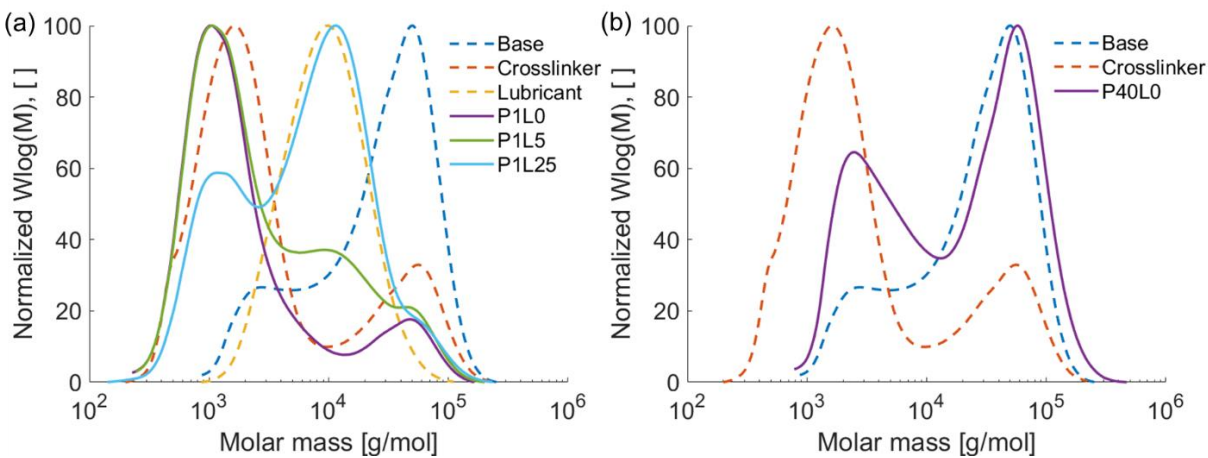

**Figure S1:** Molecular weight distribution of uncrosslinked chains in (a) P1L0, P10L5, and P10L25 and (b) P40L0 overlaid on the distributions for precursors

## 2. Determination of substrate shear modulus using micro-indention tests

Figure S2 shows a schematic of the micro-indentation test used to measure substrate Young modulus. The micromechanical testing station FT-MTA02 (FemtoTools AG, Buchs ZH,

Switzerland) consists of a micro robotic manipulation system (FT-RS1002) and a universal measurement stand with a tilt-table digital microscope (FT-UMS1002). Force sensing probes (type FT-S1000) with a range of +/-1000µN and a sensitivity of 0.05 µN were used. Spheres of cubic zirconia (Sandoz Swiss precision spheres; grade 10) with a radius $R$ of 100 µm were glued to the silicon-tip of the sensing probe using a light curing adhesive (Loctite AA3494). A contact finding step (200 µm/s) with a corresponding force threshold (10µN) was performed before each indentation measurement. An indentation speed of 0.5µm/s was applied for all samples. For each sample 4 measurement points were recorded, forming an equidistant grid of 2x2 points on an area of 0.5 x 0.5 mm.

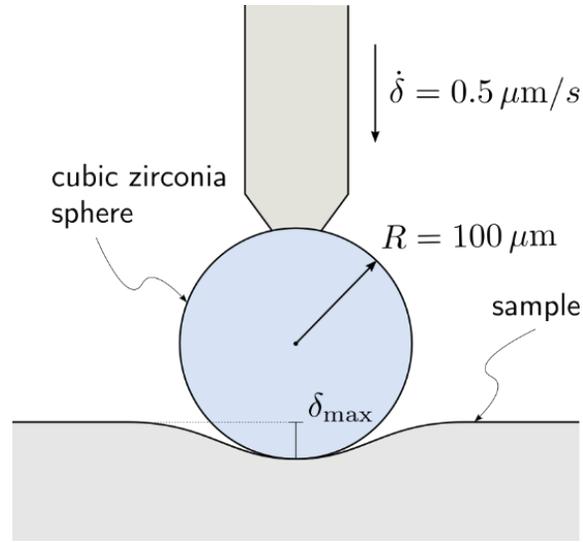

**Figure S2:** Schematic overview of the sensing probe-tip, represented in the deformed state

In order to determine the Young's modulus, a maximum indentation depth of $\delta_{max} = 5\mu m$ (Figure S1) was considered. A Hertzian contact model (1) was used, assuming a Poisson's ratio of $v=0.5$ (incompressible material) (2). A Nelder-Mead downhill simplex algorithm in Python (binaries provided by the scipy. optimize module (3)) was used to find an optimal parameter set ($\alpha$, $\beta$) of the force function $\alpha(\delta - \beta)^{\frac{3}{2}}$. The Young's modulus is then given by $E = \alpha\left(\frac{3}{4R^{\frac{1}{2}}}(1 - v^2)\right)$. For each sample the Young's modulus was calculated for all four locations. Subsequently these values were averaged over each sample. Substrate shear modulus was then calculated as $G = E/3$ (4).

## 3. Additional data on wetting ridge height

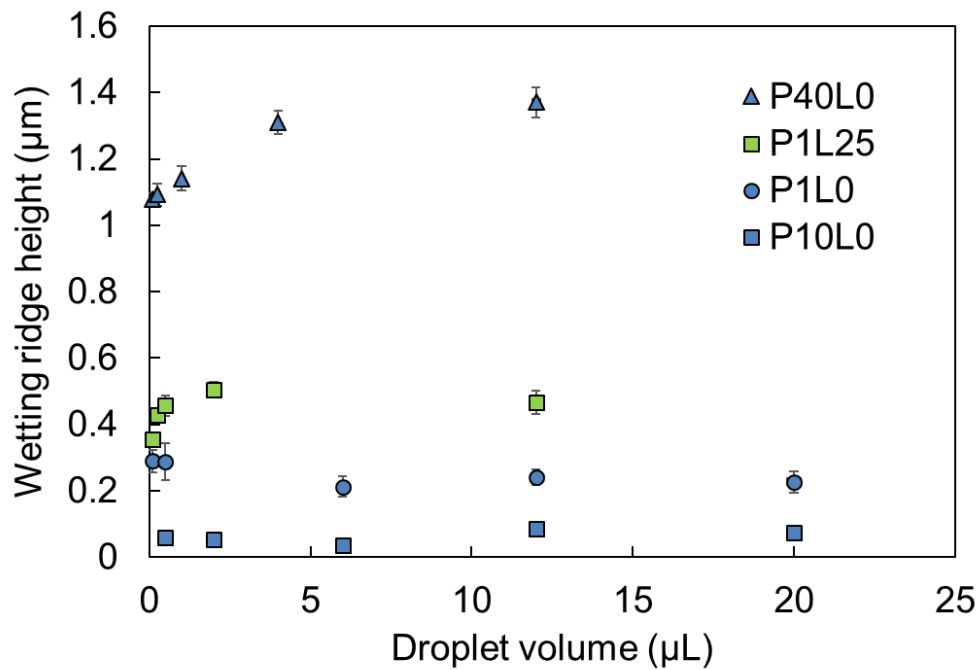

**Figure S3:** Wetting ridge height as a function of droplet volume for four substrates. The wetting ridge height shows minimal variation for the range of droplet volumes for each substrate. However, for consistency, wetting ridge heights measured using droplets of volume ~12µL or larger were reported in Figure 1 in main text.

## 4. Droplet sliding experiments

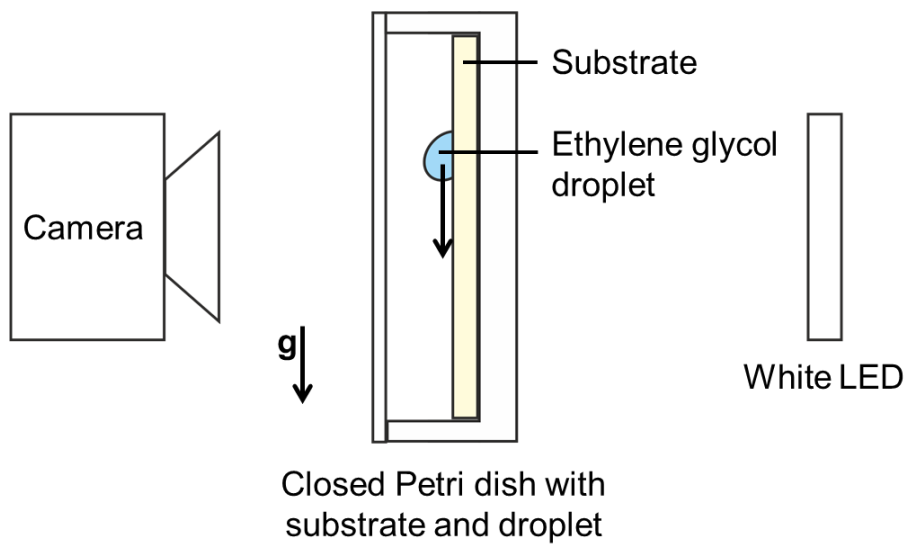

**Figure S4:** Schematic of the droplet sliding experiment

Figure S4 shows a schematic of the custom made droplet sliding experiment setup which was used to characterize the substrate viscoelasticity. The sample was contained in a Petri dish and illuminated from behind with a white LED. The cured sample was ~2 mm thick. For each measurement, individual droplets of ethylene glycol were carefully deposited on the substrate using a micropipette (Eppendorf Research) and the droplet sliding motion was observed using scientific CMOS camera. For each droplet size, the experiment was repeated for 3 droplets and two samples of each type. The values for the velocities were computed by dividing the travelled distance of droplet centroid over the corresponding time.

## 5. Microscale condensation setup

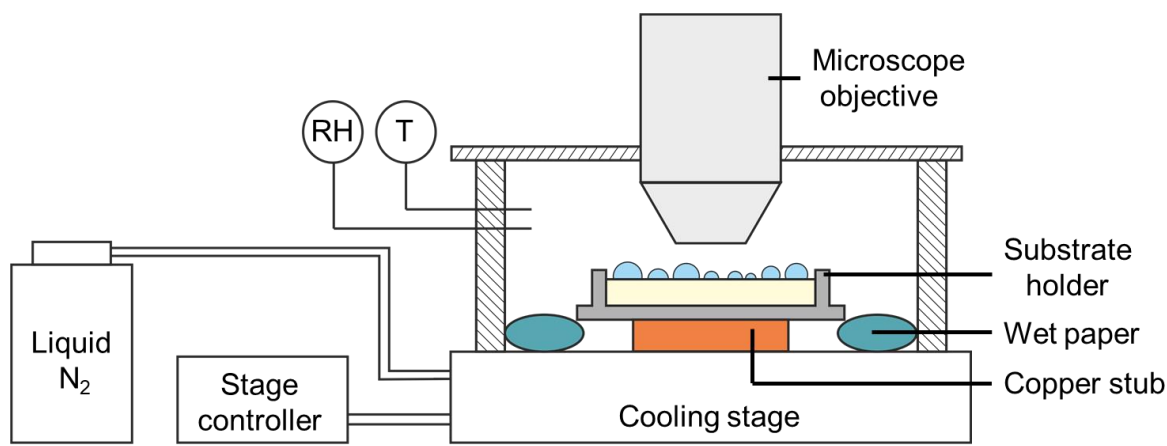

**Figure S5:** Schematic of micro-condensation setup

## 6. Dewing experiments

Figure S6 shows a schematic of the custom-built experimental setup for dew collection. It consists of a poly (methyl methacrylate) (PMMA) chamber with interior dimensions 250 mm × 180 mm × 400 mm. A 4 mm inlet at the corner of the box allows the entry of humid air at a mean flow rate of 413.2 L/h (FAM3255, ABB), generated by passing compressed air through a bubbler. A copper plate, which is screwed onto a temperature-controlled Peltier stage (CP-200TT, TE TECHNOLOGY), allows three samples to be mounted and tested simultaneously. An RTD is attached onto the copper plate to monitor its temperature. Two temperature and relative humidity sensors (SHT31, Sensirion) are placed in the box near the samples. A PMMA plate is situated between the humid air inlet and the 3 samples to reduce the effects from vortices due to the air jet. A DSLR (D7500 and AF-S DX Zoom-Nikkor 12-24mm f/4G IF-ED, Nikon) is used to monitor the condensation behaviour by taking photographs at specified intervals using its internal timer.

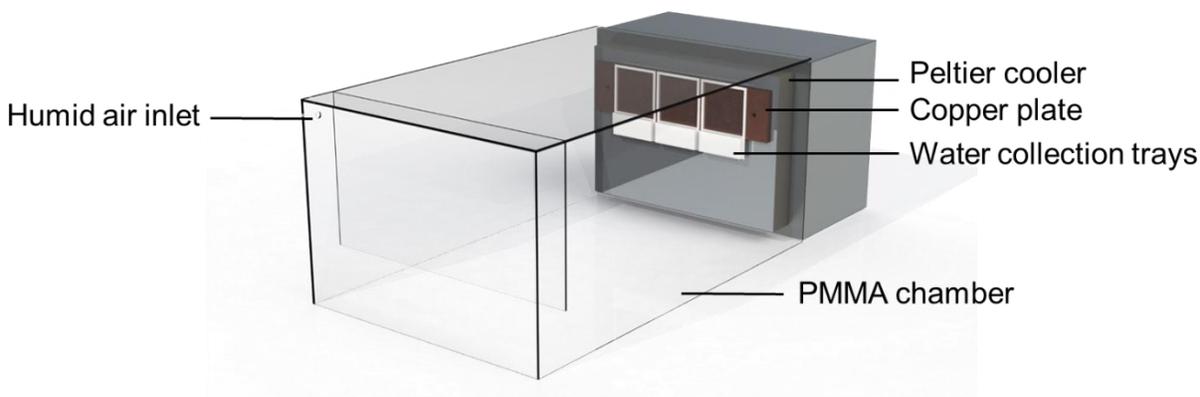

**Figure S6**: Dew collection experimental setup. A PMMA box encloses the humid air with one end open for the Peltier stage and samples. A hole is seen on the top left for incoming humid air generated by a bubbler. A plate is placed between the inlet and the samples to avoid effects from the jet affecting the flow condition at the samples. A copper plate is used to mount three samples. Dew condensed on the samples is collected in water collection trays as shown.

The 3 PDMS samples are directly fabricated on the copper plate. 3D-printed white polycarbonate mounts are attached to the copper plate. PDMS is poured into these mounts and cured in the oven with the mounts and the copper plate. The resulting cured PDMS samples of ~ 2 mm thickness thus have direct contact with the copper plate. Water condensed on the PDMS surfaces are collected in the trays of the mounts as they depart. For each PDMS sample, the dimensions of condensing area are 45 mm $\times$ 45 mm.

Experimental procedure: After the 3 PDMS samples are cured in the oven, they are allowed to cool in ambient air. The copper plate with the mounts and the samples is then screwed onto the Peltier. An RTD is attached onto the copper plate with Kapton tape. The Peltier stage is set to 25 ˚C. The temperature and humidity sensors are then placed near the samples. As humidity reaches ~ 70 %, the camera is set to capture photographs using aperture mode, at f/4 aperture and ISO 1600 at specified intervals using the internal timer, with exposure smoothing turned on. The Peltier temperature is set, resulting a copper plate temperature measured by the RTD of ~ 2 ˚C. The setup is let to run for ~ 17 hours with copper plate temperature and chamber temperature and humidity continuously monitored. After ~ 17 hours, the Peltier is switched off. The bubbler is kept running to avoid evaporation of collected water due to large drop in humidity. A cleanroom towel is used to remove water on top of the PDMS which has not departed into the tray. A surgical blade is used to remove the mount from the copper plate, with the collected water in its tray. Another cleanroom towel is used to remove water on the mount which is not in the tray. The mount is subsequently weighed. The weighing process is repeated for all 3 mounts. After that, the mounts are completely dried with nitrogen and weighed again. The difference in mass is taken as the amount of water collected for the 3 samples. Finally, the PDMS samples are removed from the copper plate and their thicknesses are measured. The experiment is repeated with the positions of the 3 samples on the copper plate shuffled to reduce effects due to the flow condition.

## 7. Energy model and dissipation ratio

As a droplet slides on the vertical surface of a soft substrate, it attains a terminal velocity $U_{drop}$ when the work done by the weight of the drop is balanced by viscous dissipation within the drop $(\dot{D}_v)$ and viscoelastic dissipation within the substrate $(\dot{D}_{ve})$. Hence $mgU_{drop} = \dot{D}_v + \dot{D}_{ve}$. Here $\dot{D}_v = \frac{6\pi\mu r l U_{drop}^2}{\theta}$ and $\dot{D}_{ve} = \frac{2r\gamma^2 I}{\pi G h}\left(\frac{U_{drop}}{U_0^m}\right)^{m+1}$ (4). Here, $\gamma$ and $\mu$ are density and viscosity of the liquid respectively, $\theta$ is the apparent contact angle and $U_0$ and $m$ are rheological constants for a substrate. $I$ and $l$ are constants based on droplet morphology with $I = \int_0^{\pi/2} \cos^{(m+1)}(\phi)\, d\phi$ and $l \approx \log(1000)$, $r$ is the base radius of the drop and $h$ is the wetting ridge height (4). The volume of the drop $V_{drop}$ is calculated as $V_{drop} = r^3 f(\theta)$ with $f(\theta) = \pi(2 + \cos\theta)(1 - \cos\theta)/[3\sin\theta(1 + \cos\theta)]$. Assuming $\dot{D}_{ve} \gg \dot{D}_v$ results in the equation $\log(U_{drop}) = \frac{1}{m}\log(V_{drop}^{2/3}) - \frac{1}{m}\log\left(\frac{2\gamma^2 I}{f^{1/3}(\theta)\pi G h U_0^m \rho g}\right)$. Fitting the experimental data to this linear relationship between $\log(U_{drop})$ and $\log(V_{drop}^{2/3})$ yields values for $m$ and $U_0$ for a substrate. With these values known, the model fit shown in Figure 2 of main text is given by $U_{drop} = \frac{U_0}{V_0^{2/3m}} V_{drop}^{2/3m}$ where $V_0 = \left(\frac{2\gamma^2 I}{\pi\rho G h f^{\frac{1}{3}}(\theta)}\right)^{3/2}$. Additionally, the dissipation ratio $\frac{\dot{D}_{ve}}{\dot{D}_v}$ can be calculated for all substrates. Table S1 list the values of apparent contact angle, model fit parameters and dissipation ratio values for various substrates. The estimated dissipation ratios show that bulk lubricant infusion in PDMS with stoichiometric monomer–to–crosslinker ratio of 10:1 results in the most reduction in viscoelastic dissipation. Additionally, comparing P1L0, P1L5 and P1L25, we find that bulk lubricant infusion in PDMS with non-stoichiometric monomer to crosslinker ratio does not lead to any significant reduction in viscoelastic dissipation. This is also evident from comparison of droplet sliding velocities for these substrates as illustrated in Figure S7.

| Substrate | $\theta$ | $m$ | $U_0$ (μm/s) | $V_0$ (μL) | $\dfrac{\dot{D}_{ve}}{\dot{D}_v}$ | $\left(\dfrac{\dot{D}_{ve}}{\dot{D}_v}\right)_{P10L0} \Big/ \left(\dfrac{\dot{D}_{ve}}{\dot{D}_v}\right)$ |
|---|---|---|---|---|---|---|
| P10L0 | 90.1±3.5 | 0.23 | 13.38 | 3.96 | $1.2 \times 10^3$ | 1 |
| P10L5 | 86.2±0.9 | 0.26 | 155.14 | 2.83 | 42.2 | 27.9 |
| P10L25 | 81.8±1.1 | 0.26 | 2062.65 | 6.05 | 20.5 | 57.4 |
| P1L0 | 79.4±2.1 | 0.49 | 171.68 | 4.73 | 133.4 | 8.8 |
| P1L5 | 75.7±2.1 | 0.41 | 812.83 | 12.05 | 115.8 | 10.1 |
| P1L25 | 77.8±3.4 | 0.39 | 1244.67 | 11.27 | 76.1 | 15.4 |
| P40L0 | 91.6±7.3 | 0.08 | 0.99 | 10.1 | $3.1 \times 10^5$ | 0.003 |

**Table S1:** Fitting parameters and dissipation ratios for various substrates estimated based on droplet sliding experiment data. Dissipation ratios are calculated for droplet of volume 5μl for all substrates except P40L0 for which droplet of volume 8 μl (minimum droplet size for which sliding velocity on P40L0 could be recorded in our experiments) is used(4).

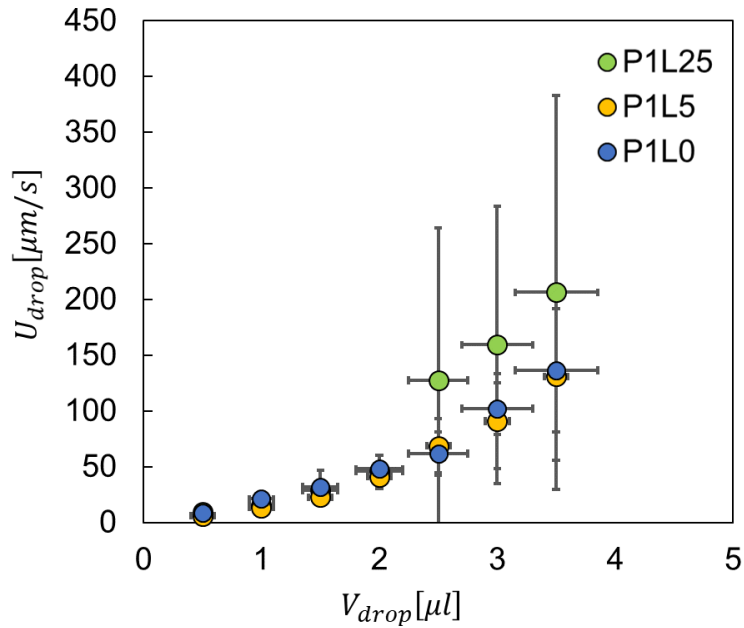

**Figure S7:** Comparison of droplet sliding velocities as a function of droplet volume for P1L0, P1L5 and P1L25

## 8. Pre-coalescence droplet growth rates on P1L0, P1L5 and P1L25

As shown in Figure S8, the comparison between pre-coalescence condensate droplet growth rates for P1L0, P1L5 and P1L25 is similar to that between P10L0, P10L5 and P10L25 shown in Fig. 4 of main text.

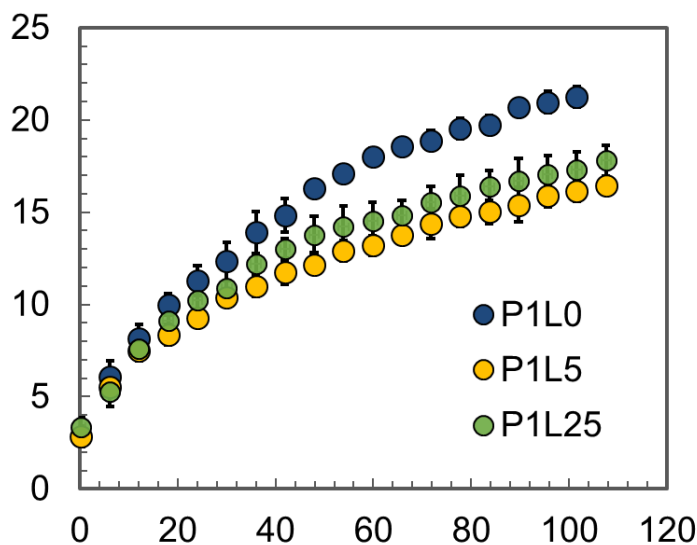

**Figure S8:** Pre-coalescence condensate droplet growth on P1L0, P1L5 and P1L25

## 9. Experimental details to measure cloaking timescale

The experimental procedure to measure differences in timescale for cloaking of water droplets between P10L0 and P10L25 consists of following steps which were automated using the Krüss Advance software.

A syringe tip (1.8 mm outer diameter) is lowered into the hole until it is a few millimetres below the lower surface. 25 μL of MilliQ water is dosed at 3 μL s$^{-1}$. Subsequently, we wait for 10 s for any vibrations from the dosing to subside and then measure the surface tension while drop is hanging from the syringe tip. This is to ensure that the initial calculated surface tension of the drop, before it comes into contact with the surface, corresponds to the expected value of ≈72 mN m$^{-1}$. Then the syringe is moved upwards until the drop touches the surface and the drop volume is increased by 13 μL to 38 μL at 5 μL s$^{-1}$. 38 μL corresponds to the maximum drop volume that can be supported by a 2 mm diameter contact line, assuming that the effective surface tension of the drop decreases to ≈60 mN m$^{-1}$ after cloaking. The surface tension measurements are taken at a frequency of 1-10 Hz, depending on how fast the process is expected to be. At the end of the experiment, the syringe tip is pulled out of the drop and the surface tension of the pendant drop is measured (without the tip attached to it). This is to check if the presence of the syringe tip was influencing the measured surface tension.

Repeated measurements on the same substrate shows variations in the cloaking timescale. However, cloaking always happened significant faster on P10L25.

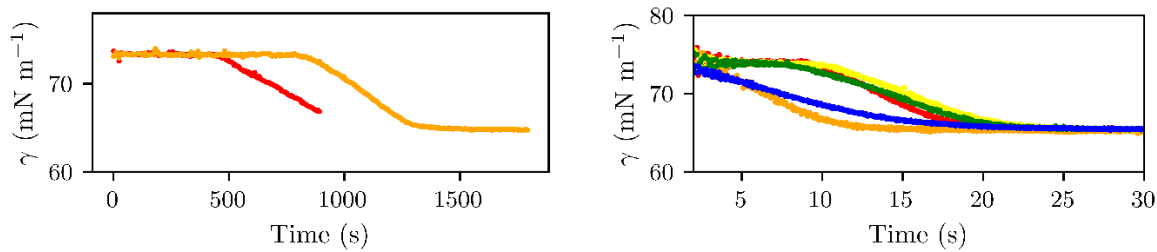

**Figure S9:** Evolution of surface tension of a water drop hanging from P10L5 (left) and P10L25 (right). The different colors on each plot correspond to repeats of the same experiment. The order of the repeats corresponds to the rainbow colors, with red corresponding to the first experiment. The time between each repeat experiment was ≈ 5 minutes. With P10L0, there is no change the measured surface tension for over 30 minutes.

## 10. Supplementary video

Supplementary video shows condensation on P10L0, P10L5 and P10L25 under an atmospheric relative humidity, temperature of humid air and surface temperature ~70% and ~ 17℃ respectively and copper plate temperature of ~2℃.

## References

1.  Johnson KL (1985) *Contact Mechanics* (Cambridge University Press, Cambridge) doi:DOI: 10.1017/CBO9781139171731.

2.  Hopf R, et al. (2016) Experimental and theoretical analyses of the age-dependent large-strain behavior of Sylgard 184 (10:1) silicone elastomer. *J Mech Behav Biomed Mater* 60:425–437.

3.  Jones E, Trevor O, Pearu P Scipy: Open Source Scientific Tools for Python.

4.  Shanahan MER, Carré A (2002) Spreading and dynamics of liquid drops involving nanometric deformations on soft substrates. *Colloids Surfaces A Physicochem Eng Asp* 206(1–3):115–123.